\documentclass[12pt]{iopart}

\usepackage{setspace} 
\usepackage{color}

\newcommand\JMC[1]{\textcolor{black}{  #1}}

\usepackage{graphicx}

\begin{document}

\title{Identification of redundant and synergetic circuits in triplets of electrophysiological data}
\author{Asier  Erramuzpe$^{1}$, 
Guillermo J. Ortega$^{2}$, 
Jesus Pastor$^{2}$,
 Rafael G.  de Sola$^{2}$,
Daniele Marinazzo$^{3}$,
Sebastiano Stramaglia$^{1,4,5}$\footnote{New address: BCAM, The Basque Center for Applied Mathematics. Bilbao, Spain }, and Jesus M. Cortes$^{1,4,6}$}
\address{$ˆ1$  Biocruces Health Research Institute. Cruces University Hospital,   Barakaldo, Spain }
\address{$ˆ2$ Instituto de Investigacion Sanitaria Hospital de la Princesa, Madrid, Spain}
\address{$ˆ3$ Department of Data Analysis at the  Faculty of Psychology and Educational Sciences. Ghent University, Belgium}
\address{$ˆ4$ Ikerbasque: The Basque  Foundation for Science. Bilbao, Spain}
\address{$^5$ Dipartimento di Fisica, Universita degli Studi di Bari and INFN. Bari, Italy}
\address{$^6$ Department of Cell Biology and Histology. University of the Basque Country. Leioa, Spain}
\ead{jesus.cortesdiaz@osakidetza.eus}

\clearpage
\begin{abstract}

\JMC{\textit{Objective}. Neural systems are comprised of interacting units, and relevant information regarding their function or malfunction can be inferred by analyzing the statistical dependencies between the activity of each unit. Whilst correlations and mutual information are commonly used to characterize these dependencies, our objective  here is to extend interactions to triplets of variables to better detect and characterize dynamic information transfer.
 \textit{Approach}. Our approach relies on the measure of interaction information (II). The  sign of II provides information as to the extent to which the interaction of variables in triplets  is redundant (R) or synergetic (S). 
Three variables are said to be redundant when a third variable, say Z, added to a pair of variables (X,Y), diminishes the information shared between X and Y. Similarly, the interaction in the triplet is said to be synergetic  when   conditioning on  Z enhances the information shared between X and Y with respect to the unconditioned state. Here, based on this approach, we calculated the R and S status for triplets of electrophysiological data recorded from drug-resistant patients with mesial temporal lobe epilepsy in order to study   the spatial organization and  dynamics of R and S close to the epileptogenic zone (the area responsible for seizure propagation).
\textit{Main results.} 
In terms of   spatial organization, our results show that R matched the epileptogenic zone while S was distributed more in the surrounding area. In relation to  dynamics, R made the largest contribution to high frequency bands (14-100Hz), whilst S was expressed more strongly at lower frequencies (1-7Hz). Thus, applying interaction information to such clinical data reveals new aspects of epileptogenic structure in terms of the nature (redundancy vs. synergy) and dynamics (fast vs. slow rhythms) of the interactions. \textit{Significance.} We expect this methodology, robust and simple,  can reveal new aspects beyond pair-interactions in networks of interacting units in other setups with  multi-recording data sets (and thus, not necessarily in epilepsy, the pathology we have approached here).}

\end{abstract}

\clearpage

\section*{Introduction}

The use of information theory to deal with data has become an important means to evaluate the interaction between groups of correlated variables in neuroscience, revealing functional relationships and the underlying circuits capable of processing \JMC{specific} information \cite{borst1999,panzeri1999,quiroga2009,cortes2014}.    
In addition to information storage/coding/decoding, information theory can address whether interactions between related variables are mutually redundant or synergetic  \cite{schneidman2003,bettencourt2007}.

In general, synergy (S) occurs if the understanding of one variable \JMC{helps} predict the behaviour of another variable with more precision than the sum of the information provided individually by the two variables. By contrast, redundancy (R) corresponds to situations where the same information is offered by the variables (see also the interpretation of S and R based on causality inference in \cite{stramaglia2014}).

When  interaction information (II) \cite{mcgill1954} is applied to sets of three variables, its sign \JMC{indicates} whether the relationship \JMC{within the} triplet can be considered as R or S. Thus, unlike mutual information, II can be either positive or negative, whereby a positive II identifies R and a negative II S. As an example of the mechanisms that produce both R and S, common-cause structures (such a common-input) lead to R, while S could be produced   by \JMC{combining} one XOR gate with two independent random inputs \JMC{(for instance)} \cite{lizier2012,wibral2014}.

The presence of S is well-known in sociological \JMC{modelling}, where the \JMC{term}
{\it suppressors}   \JMC{was applied} to those variables \JMC{that increase}   the predictive validity of other variables \JMC{after their  inclusion in}  a linear regression equation \cite{conger1974}. Similarly, the interaction between triplets of variables in gene regulatory networks was approached to study how a specific gene modulates the interaction between two other genes \cite{antonov2004,wang2009}. Moreover, when applied to electrophysiological recordings from neural networks in culture,   a series expansion approach takes into account terms that are coincident with the II \cite{bettencourt2008}. Here, we use II to \JMC{study} triplet interactions in epilepsy but rather than focusing on the pathology, we aim to highlight \JMC{here} the methodological aspects.   \JMC{In particular, by applying    II  to human electrocorticography data, we   reveal  certain aspects of the interactions (not captured by looking solely  to pairs of variables) that underpin the epileptogenic zone, the brain network that triggers epileptic seizures.}

\section*{Methods}

\subsection*{Surgery and postsurgical outcome}

The selection of the area for resection was made according to current standard practices that are applied to surgery for epilepsy, which involves resecting the cortical area \JMC{that displays the most excitatory activity  electrophysiologically } as a proxy of the epileptogenic zone \cite{rosenow2001}. 

The clinical outcome was assessed \JMC{using} the Engel Epilepsy Surgery Outcome Scale \cite{engel1993}, ranging from I (seizure-free after surgery) to IV (no improvement after surgery). The electrocorticography data studied here correspond to n=3 Engel I   patients that had no further seizures after surgery and n=1  Engel III patient   for comparison.

\subsection*{Human electrocorticography data}
The data analyzed here through the novel II approach has been published previously \cite{palmigiano2012}  and \JMC{correspond} to  data recorded from drug-resistant patients with temporal lobe epilepsy who underwent surgery at the Epilepsy Unit at “La Princesa” Hospital (Madrid, Spain). After receiving the approval of the Ethics Committee at “La Princesa” Hospital, patients provided their informed consent to be evaluated intraoperatively with a $4\times5$  subdural electrode grid (interelectrode distance, 1 cm), having been administered low doses of sevoflurane (0.5\%) and remifentanil (0.1 mg/kg/min). 
While recording, the anaesthesia was stabilized within a bispectral index in the range of  55-60 (adimensional index)   \cite{rosow2001}.
The grid was placed over the lateral temporal cortex, with its   border parallel to the sylvian fissure and covering gyri T1-T3, and its position was recorded with a video camera or photographed. 
 The reference electrode was placed on the contralateral ear and in some cases it was moved to the nearby scalp in order to verify there was no contamination of the recording, \JMC{consistent with previous studies}    \cite{ortega2008}.
A presurgical evaluation was carried out according to the protocol used at La Princesa Hospital, as reported previously \cite{pastor2005}.

From the \JMC{perspective of}  signal processing, it is important to emphasize the importance of \JMC{electrocorticography studies as opposed to those based on scalp electroencephalography.  In the former signal quality is much better, with a signal to noise ratio about 21 to 115 times greater, which makes these two modalities quite distinct  \cite{ball2009}.}

\subsection*{Data processing}
All of the analyses carried out here were performed retrospectively and thus, tailored  \JMC{lobectomies were}  not based on the results discussed here. An intraoperative electrocorticography session was recorded for 15-20 min using a 32-channel amplifier (Easy EEG II, Cadwell, USA), preprocessing with a filter with a 0.5-400 Hz bandwidth and with a 50 Hz notch filter, and finally downsampling at 200 Hz. Artefact-free epochs of inter-ictal activity lasting up to 5 min were selected by visual inspection. All post-processing analysis was performed in Matlab (MathWorks Inc., Natick, MA).

\subsection*{Frequency-band analysis}
Electrophysiological signals were filtered within different frequency bands using a zero-phase digital filter (figure  \ref{fig3}). In particular, we used the   \textit{filtfilt.m} function from Matlab (MathWorks Inc., Natick, MA) to filter  \JMC{the following standard bands for brain electrophysiology}: delta (1-4 Hz), theta (4-7 Hz), alpha (7-14 Hz), beta (14-26 Hz), and gamma (26-100 Hz).

\subsection*{Shannon entropy}

The Shannon entropy of a random variable  $X$ (i.e.: its average uncertainty) is defined as

\begin{eqnarray}
H(X)=-\sum_x \mathrm{prob}(x) \mathrm{log} \,\,\mathrm{prob}(x),
\label{h}
\end{eqnarray}
 where $\mathrm{prob}(x)$ represents the probability distribution of the   state $x$ of variable $X$ \cite{jaynes1957,cover2006}.  In this manuscript, variable X is defined by the electrical potential captured by electrode X. The joint entropy is just a generalization to any dimension, i.e., in 2D it becomes  $H(X,Y)=-\sum_x \sum_y \mathrm{prob}(x,y) \mathrm{log} \,\,\mathrm{prob}(x,y) $ and  in 3D,   we have $H(X,Y,Z)=-\sum_x \sum_y    \sum_z \mathrm{prob}(x,y,z) \mathrm{log} \,\,\mathrm{prob}(x,y,z).$ For a base 2 logarithm (as used here), the entropy is expressed in bits.

\subsection*{Interaction information}

II  is   measured based on the Shannon entropy, 
 allowing  us  to analyze   \JMC{the} interactions between triplets \cite{mcgill1954}. For any triplet $(X, Y, Z)$,   II   is defined as
\begin{eqnarray}
\mathrm{II}(X,Y,Z) &\equiv &  I(X,Y)-I(X,Y|Z)
\label{ii}
\end{eqnarray}
where $ I(X,Y)$ is the mutual information between $X$ and $Y$, \JMC{which can be  defined in terms of the marginal and joint entropies,  ie.,}

\begin{eqnarray}
I(X,Y)&=&H(X)+H(Y)-H(X,Y), 
\label{mutual}
\end{eqnarray}
  \JMC{and $I(X,Y|Z)$ is the  conditional  mutual information between $X$ and $Y$ conditioned to $Z$. Analogously,    $I(X,Y|Z)$  can be written   as a function of the marginal and joint entropies, i.e.,   }

\begin{eqnarray}
I(X,Y|Z)&=&H(X,Z)-H(Z)+H(Y,Z)-H(X,Y,Z),
\label{condmutual}
\end{eqnarray}
\JMC{for further details see \cite{cover2006}. Thus, using the definition given by Eq. (\ref{ii}), and  Eqs.  (\ref{mutual}) and (\ref{condmutual}), one can express II as a function of the Shannon entropies, i.e., }

\begin{eqnarray}
\mathrm{II}(X,Y,Z) &\equiv& H(X,Y,Z)+H(X)+H(Y)+H(Z)   \nonumber \\
&& -H(X,Y)-H(X,Z)-H(Y,Z). 
\label{ii_shannon}
\end{eqnarray}
\JMC{From this equation, one can derive  simple but important properties of II (for details see Appendix A).}

Note that our definition of II uses an opposite sign to the original form in \cite{mcgill1954}. Also, it is important to emphasize that when II is equal to zero, II is ill-posed as two different situations can correspond to II = 0, namely that the three variables are either statistically independent of each other or that  $I(X, Y) \approx I(X, Y|Z)$. \JMC{Therefore, we will only report   situations satisfying that  II$ \,\neq0$, i.e., values of  II$\,\approx0$ will   be ignored.}

 \subsection*{Calculation of the interaction information assuming Gaussian data}

We calculated the II under a Gaussian approximation for which the Shannon entropy can be analytically calculated, i.e.: where the differential entropy for the multivariate Gaussian distribution has an analytical derivation (see details in \cite{cover2006}).
To calculate the conditional mutual information, Eq. (\ref{condmutual}),  we made use of \cite{barnett2009}, where for  multivariate Gaussian random variables it was shown that:

\begin{eqnarray}
I_{\mathrm{gaussian}}\left(X;Y|Z\right)=\frac{1}{2}\ln{|\Sigma(X|Z)|\over|\Sigma(X|Y\oplus
Z)|},
\label{gci}
\end{eqnarray}
where $|\cdot|$ denotes the determinant, and the partial covariance
matrix is defined by
\begin{eqnarray}
\Sigma(X|Z)=\Sigma(X)-\Sigma(X,Z)\Sigma(Z)^{-1}\Sigma(X,Z)^\top,
\label{gcov}
\end{eqnarray}
in terms of the covariance matrix $\Sigma(X)$ and the cross
covariance matrix $\Sigma(X,Z)$; the definition of $\Sigma(X|Y\oplus
Z)$ is analogous, where $Y\oplus Z$ means appending the two variables.

To calculate II, Eq. (\ref{ii}),  we also need to calculate the mutual information term; obtained using the same Eq.  (\ref{gci})   but in absence of $Z$.

\subsection*{Calculation of the interaction information assuming non-Gaussian data}
In the most general situation of data not following a Gaussian distribution, and in order to calculate the mutual information and the conditional mutual information (cf. rhs in Eq. (\ref{ii}), \JMC{we first binned the data to   4 states 
and then evaluated equations  (\ref{mutual}) and (\ref{condmutual})  using the  Mutual Information Toolbox \cite{peng2005}, a C++  implementation of several Information Theory based functions  plugged in  Matlab  (MathWorks Inc., Natick, MA).  In particular, we made use of the functions \textit{mutualinfo.m} and \textit{condmutualinfo.m} incorporated in \cite{peng2005}.  }   To calculate probabilities, also 6 and 10 bins were   used and the results were no different.

\subsection*{Redundancy and synergy}

Directly from the definition of II, Eq. (\ref{ii}), a positive sign of II means that $I(X,Y) > I(X,Y|Z)$, and together with the circulation property (establishing that II is invariant to any possible permutation in $(X, Y, Z)$; for further details see Appendix A)  this implies that the interaction in the triplet is redundant. Similarly, a negative sign of II corresponds to \JMC{a} synergetic interaction.

This is in agreement with the results based on \JMC{the} partial information decomposition reported in \cite{williams2010nonnegative}, where a similar interpretation of the sign of II was provided: a positive one for R and a negative one for S, taking into account the fact that the right hand side of Eq. (\ref{ii}) has the opposite sign to the II reported in \cite{williams2010nonnegative} \JMC{(see also a further discussion in \cite{barrett2014})}.

\subsection*{Statistical significance}

All values reported in this manuscript were statistical significant. Statistical significance was approached by building the null-distribution of no interaction after a shuffling procedure of N=50000 repetitions in the time series used for calculation. Significant values (after Bonferroni correction) were obtained with a p-value of 0.05 for both the S and R tails of the distribution (negative and positive values), as II is not normally distributed,   \JMC{i.e., the tail of the II distribution has a different  length on the positive and negative sides.}

\subsection*{Averaging of the interaction information} 

For each of the 20 grid electrodes, we used different non-overlapping windows, ranging from 9 to 29 windows, each containing 2,000 time points (10 seconds at a sampling rate of 200 Hz). The values of II reported here correspond to the average II over all the different windows, producing an appropriate sample  (figure \ref{figS1}).

\subsection*{Network measures for plotting interaction information in 2D}

To obtain 2D plots of II, whereby each default value depends on a triplet of variables, we represented the II variation across two variables while keeping a third variable fixed. 

 \JMC{In order to study the network of interactions rather than looking to individual values, we considered   each variable to be a node within a network  and after   applied different network measures using the Brain Connectivity Toolbox \cite{rubinov2010,bctURL}.
In particular, the fixed electrode was chosen as the one with  maximum betweenness centrality (accounting for the fraction of all the shortest paths in the network that contain a given node, and thus,   nodes with high betweenness   participate in many of the shortest paths), that with maximum degree (the hub, obtained simply by summing   the weighted links connected to a given node),  and the one with a maximum   clustering coefficient  (calculated as the fraction of triangles around a node, and thus providing relevant information of the clustering strength in a node neighbourhood). 
}

\subsection*{Local synchronization index, individual redundancy and individual synergy}

In the light of previous studies based on the local synchronization index (LSI) in electrocorticography data  \cite{palmigiano2012}, we were interested in  indexes that can work \JMC{for} individual electrodes.    As reported in \cite{palmigiano2012}, \JMC{the LSI measures the average synchronization of each electrode with their first neighbours, including those in the  diagonals. In particular, LSI is calculated by averaging  the absolute value of the Pearson correlation between a given node and its nearest neighbours,   i.e., 3 neighbours at corner's grid, 5 neighbours at side's grid and 8 neighbours elsewhere. }

\JMC{In an analogous manner, the individual R was calculated for a given electrode   $\bar{Z}$, summing all the values of positive     $II(X,Y,\bar{Z})$ for all  $X$ and $Y$  variables. Likewise, the individual  S was obtained by summing for all    $X$ and $Y$ the negative  values of     $II(X,Y,\bar{Z})$ and  finally taking the absolute value of this sum. For these two measures, the individual R and S values, we only summed on triplets such that the three variables were distinct to one another.}

\section*{Results}

\JMC{II depends on triplets of variables and its positivity or negativity  identifies rather the  interaction in the triplet is (respectively) redundant or synergetic.  First, we inspected the patterns of II by projecting it to the 2D plots,    fixing  one of the electrodes  and plotting II by varying the other two   variables   (see Methods).   }

\JMC{We  first fixed  the electrodes with maximum and minimum entropy, two  good references for studying  II as II(X,X,X) coincides with the entropy (see Appendix A). We also fixed several electrodes that at the network level play an important role such as the one    with maximum betweenness, the one with  maximum degree  of connectivity (i.e.: the hub) and the one with a maximum clustering coefficient (see Methods for further details on  these measures). As it is shown in  figure \ref{fig1},   the 2D plots  drastically depend on what the   electrode  was fixed, indicating a clear 3D structure of II, i.e., a very distinct II patterns  appears in the neighbourhood of the different electrodes.}

\JMC{Next, we assessed whether the    R and S patterns  were  in any way related with the epileptogenic zone (the area responsible for seizure propagation). This could be addressed as some of the patients had an  Engel I outcome (Methods) and did not experience further seizures after surgery. Given the resection area (an area known for all patients) and using the recordings before surgery, it is possible to guarantee  that the resection area contained at least part of the epileptogenic zone. In addition to the resection area (the gray grid in figure \ref{fig2}), we also calculated the spatial distribution of S (red grid in figure \ref{fig2}), R (blue grid),  Shannon entropy (green grid) and  LSI (the magenta grid), an index that was proven to match the putative epileptogenic zone \cite{palmigiano2012}.  Accordingly, while R and  LSI had the same spatial grid distribution, the organization of S surrounded them,  an observation  that was even clearer  when   II was calculated under the assumption of non-Gaussian data (figure \ref{fig6}). The pattern described by the Shannon entropy was more dispersed, and thus, less related to the epileptic zone.}

\JMC{Next, we asked whether the relationship between S/R and LSI  (a proxy of the epileptogenic zone)  preserved across different brain rhythms. 
 Figure    \ref{fig3}  shows  2D plots of    II   across different frequency bands,    keeping fixed one of the electrodes corresponding to the resection area.
From top to bottom in figure \ref{fig3}, the lower the panel, the fastest the rhythm was. Thus, straightforward from figure \ref{fig3}, II increased for faster  rhythms   --  beta (14-26 Hz) and gamma (26-100 Hz)  bands, the two lowest panels -- in comparison to slower ones.    }

Regarding R and S across   frequency bands (figure  \ref{fig4}), R had a stronger contribution to high-frequency beta and gamma bands (in orange and brown), while both bands showed a similar spatial  distribution, coincident with the LSI that is known to match the epileptogenic zone. \JMC{By contrast,} the low delta and theta bands had a stronger contribution to S (dark and light blues). Importantly, the spatial distribution of S differed considerably \JMC{from} that of R, and while the activation of R matched the epileptogenic zone  (i.e., LSI)   at high frequencies, S surrounded it and contributed more  at low frequency bands.

In terms of the robustness of our conclusions across subjects, figures \ref{fig1}-\ref{fig4}  are data from one of the patients with Engel I, but similar results were observed for the other patients with the same Engel I outcome. 
Indeed, figure \ref{fig5} shows the total amount of R inside the resection area and the total S outside the resection area. For Engel I patients, where the resection area was accurate and patients \JMC{suffered} no further crises, the patterns identified were robust, unlike those from the Engel III patient. This is better captured when the ratio of these quantities   \JMC{ is considered, with R within the resection area and S organized across areas surrounding it (see figure 5). In addition, the total R of the Engel III patient is much higher than that of the other three Engel I patients.}

\section*{Discussion}

\JMC{Studying the interactions among variables in triplets, such as that captured by II, goes beyond bivariate and multivariate studies of interacting variables. Although this method has previously been applied to other fields, to the best of our knowledge we are unaware of similar studies in epilepsy. We have continued earlier studies that took advantage of Granger causality  \cite{granger1969} to assess the effects of S and R on epileptic data \cite{marinazzo2010,wu2011,stramaglia2014}, and we calculated the II in electrocorticography data from epileptic patients who underwent surgery. In particular, we paid special attention to the data from patients that achieved the best possible outcome following surgery and who experienced no further seizures (Engel I patients). Indeed, by taking the resection area in these patients as a reference, we addressed the patterns of II near to the epileptogenic zone, the area essential for seizures to propagate.}

\JMC{We calculated the II in two situations, assuming that the data adopts either a Gaussian or a non-Gaussian distribution. While the former permits an analytical derivation of the II, the non-Gaussian distribution is more general and it is not easy to choose the most appropriate method to estimate probabilities. We are aware that we have applied the simplest method to generate histograms (binning and counting frequencies), as the number of data points was considered to be sufficiently large to guarantee an appropriate sample \cite{bonachela2008}. Nevertheless, other methods could also be used to estimate probability density, such as kernel estimators, the nearest neighbour method, orthogonal series estimators, etc. (see \cite{bishop2006} for further details). }

\JMC{Nevertheless, by assuming Gaussian and non-Gaussian distributions we obtained qualitatively similar results. That is, although  data are non-Gaussian across segments, electrodes and patients (figure \ref{figS3}, in addition of obtaining p values of zero after a Kolmogorov-Smirnov for normality test), the Gaussian assumption captures the non-Gaussian character of II quite well. Indeed, although some differences were found with respect to S, the same pattern was observed for R. Importantly, the non-Gaussian assumption was consistent with the Gaussian one in situating S outside the R area.}

\JMC{ It is important to emphasize that our aim was to provide a more complete characterization of the informational pattern in the neighbourhood of the epileptogenic zone. Accordingly, the LSI has already been shown to satisfactorily localize the epileptogenic zone \cite{palmigiano2012}. What we found here was that R matched the LSI while S was evident in its surroundings for patients with Engel I.  This pattern was particularly clear for II when assuming a non-Gaussian distribution, which is in accordance with the organization of local intracortical interactions that, under the influence of local inhibitory circuits, are responsible for controlling runway excitation \cite{ortega2008}. The matching between LSI and R was less clear for a patient with Engel III (figure \ref{figS2}). }

\JMC{Finally, we studied II (including S and R) in relation to interictal activity. However, recent results suggest it might be of interest to carry out further studies into the application of II to pre- and post-ictal activity  \cite{mierlo2014}, thereby assessing how the structure of R and S changes with ongoing seizure dynamics.  }

\section*{Acknowledgments}
The authors acknowledges Prof. Dante R. Chialvo for having introduced Guillermo J. Ortega to the Computational Neuroimaging Group at Biocruces, making this collaboration possible. We also acknowledge financial support from Ikerbasque: The Basque Foundation for Science, Gobierno Vasco (Saiotek SAIO13-PE13BF001), Junta de Andalucia (P09-FQM-4682) and Euskampus at UPV/EHU to JMC; Ikerbasque Visiting Professor at Biocruces and project “BRAhMS – Brain Aura Mathematical Simulation” (AYD-000-285), co-funded by Bizkaia Talent and European Commission through COFUND programme to SS; pre-doctoral contract from the Basque Government, Eusko Jaurlaritza, grant PRE/2014/1/252, to AE.

\clearpage

\section*{References}

\begin{thebibliography}{10}
\expandafter\ifx\csname url\endcsname\relax
  \def\url#1{{\tt #1}}\fi
\expandafter\ifx\csname urlprefix\endcsname\relax\def\urlprefix{URL }\fi
\providecommand{\eprint}[2][]{\url{#2}}

\bibitem{borst1999}
Borst A and Theunissen F 1999 {\em Nat. Neurosci.\/} {\bf 2} 947--967

\bibitem{panzeri1999}
Panzeri S, Schultz S, Treves A and Rolls E 1999 {\em Proc. Bio.l Sci.\/} {\bf
  266} 1001--1012

\bibitem{quiroga2009}
Quiroga R and Panzeri S 2009 {\em Nat. Rev. Neurosci.\/} {\bf 10} 173--185

\bibitem{cortes2014}
Cortes J, Marinazzo D and Munoz M 2014 {\em Front. Neuroinform.\/} {\bf 8} 86

\bibitem{schneidman2003}
Schneidman E, Bialek W and Berry M 2003 {\em J. Neurosci.\/} {\bf 23}
  11539--11553

\bibitem{bettencourt2007}
Bettencourt L, Stephens G, Ham M and Gross G 2007 {\em Phys. Rev. E\/} {\bf 75}
  021915

\bibitem{stramaglia2014}
Stramaglia S, Cortes J and Marinazzo D 2014 {\em New J. Phys.\/} {\bf 16}
  105003

\bibitem{mcgill1954}
McGill W 1954 {\em Psychometrika\/} {\bf 19} 97--116

\bibitem{lizier2012}
Lizier T, Atay F and Jost J 2012 {\em Phys Rev E\/} {\bf 86} 026110

\bibitem{wibral2014}
Wibral M, Lizier J, Voegler S, Priesemann V and Galuske R 2014 {\em Front.
  Neuroinform.\/} {\bf 8} 1

\bibitem{conger1974}
Conger A 1974 {\em Educational and Psychological Measurement\/} {\bf 34} 35--46

\bibitem{antonov2004}
Antonov A, Tetko I, Mader M, Budczies J and Mewes H 2004 {\em Bioinformatics\/}
  {\bf 20} 644--652

\bibitem{wang2009}
Wang K, Saito M, Bisikirska B, Alvarez M, Lim W, Rajbhandari P, Shen Q,
  Nemenman L, Basso I, Margolin A, Klein U, Dalla-Favera R and Califano A 2009
  {\em Nat. Biotechnol.\/} {\bf 27} 829--839

\bibitem{bettencourt2008}
Bettencourt L, Gintautas V and Ham M 2008 {\em Phys. Rev. Lett.\/} {\bf 100}
  238701

\bibitem{rosenow2001}
Rosenow F and Luders H 2001 {\em Brain\/} {\bf 124} 1683--1700

\bibitem{engel1993}
Engel J, Ness P~V, Rasmussen T and Ojemann L 1993 {\em Surgical Treatment of
  the Epilepsies\/} (New York: Raven Press) chap Outcome with respect to
  epileptic seizures

\bibitem{palmigiano2012}
Palmigiano A, Pastor J, de~Sola R~G and Ortega G 2012 {\em PLoS One\/} {\bf 7}
  e41799

\bibitem{rosow2001}
Rosow C and Manberg P 2001 {\em Anesthesiol. Clin. North America\/} {\bf 19}
  947--966

\bibitem{ortega2008}
Ortega G, de~la Prida L~M, de~Sola R~G and Pastor J 2008 {\em Epilepsia\/} {\bf
  49} 269--280

\bibitem{pastor2005}
Pastor J, Hernando-Requejo V, Dominguez-Gadea L, de~Llano I, Meilan-Paz M,
  Martinez-Chacon J and de~Sola R~G 2005 {\em Revista Neurologia\/} {\bf 41}
  709--716

\bibitem{ball2009}
Ball T, Kerna M, Mutschler I, Aertsen A and Schulze-Bonhage A 2009 {\em
  Neuroimage\/} {\bf 46} 708--716

\bibitem{jaynes1957}
Jaynes E 1957 {\em Phys. Rev. 106\/} {\bf 106} 620–630

\bibitem{cover2006}
Cover T and Thomas J 2006 {\em Elements of {I}nformation {T}heory\/} (John
  Wiley \& Sons, Inc)

\bibitem{barnett2009}
Barnett L, Barrett A and Seth A 2009 {\em Phys. Rev. Lett.\/} {\bf 103} 238701

\bibitem{peng2005}
Peng H, Long F and Ding C 2005 {\em IEEE Trans. Pattern Anal. Mach. Intell.\/}
  {\bf 27} 1226–1238

\bibitem{williams2010nonnegative}
Williams P~L and Beer R~D 2010 {\em arXiv preprint arXiv:1004.2515\/}

\bibitem{barrett2014}
Barrett A 2015 {\em Phys. Rev. E\/} {\bf 91} 052802

\bibitem{rubinov2010}
Rubinov M and Sporns O 2010 {\em Neuroimage\/} {\bf 52} 1059--1069

\bibitem{bctURL}
Brain {C}onnectivity {T}oolbox \url{https://sites.google.com/site/bctnet/}

\bibitem{granger1969}
Granger C 1969 {\em Econometrica\/} {\bf 37} 424--438

\bibitem{marinazzo2010}
Marinazzo D, Liao W, Pellicoro M and Stramaglia S 2010 {\em Phys. Lett. A\/}
  {\bf 374} 4040--4044

\bibitem{wu2011}
Wu G, Chen F, Kang D, Zhang X, Marinazzo D and HChen 2011 {\em IEEE Trans.
  Biomed. Eng.\/} {\bf 58} 3088--3096

\bibitem{bonachela2008}
Bonachela J, Hinrichsen H and Munoz M 2008 {\em J. Phys. A\/} {\bf 41} 202001

\bibitem{bishop2006}
Bishop C 2006 {\em Pattern recognition and machine learning\/} (Springer)

\bibitem{mierlo2014}
van Mierlo P, Papadopoulou M, Carrette C, Boon P, Vandenberghe S, Vonck K and
  Marinazzo D 2014 {\em Prog. Neurobiol.\/} {\bf 121} 19--35

\end{thebibliography}

\providecommand{\newblock}{}

\clearpage

\section*{Appendix A: Properties satisfied by the interaction information}

\JMC{Directly from equation (\ref{ii_shannon}),   one can derive important   properties which are  particularly relevant to the present work:}

\begin{itemize}
\item \textit{Circulation}: II  is invariant to any possible permutation in $(X, Y, Z)$ and  thus, the evaluation of II gives  the same result for   each of the following six  situations  $(X, Y, Z)$, $(X, Z, Y)$,  $(Y, X, Z)$, $(Y, Z, X)$, $(Z, X, Y)$,  $(Z, Y, X)$. \JMC{The proof is straightforward just   using the  Eq. (\ref{ii_shannon}).}

\item  \textit{Mutual information limit}: in the case of   triplets with only 2 distinct variables, i.e.:  $(X, Y, X)$, $(X, X, Y)$,  $(Y, X, X)$,   $II= I(X,Y)$ is satisfied, thus, II is equal to the mutual information between the two distinct variables. The proof of this    is straightforward as, by   definition, the entropy satisfies that $H(X,Y,Y)=H(X,Y)$ and  that $H(Y,Y)=H(Y)$. Thus, the definition of the mutual information given in Eq. (\ref{mutual}) provides the proof.

\item  \textit{Shannon entropy limit}: in the case of triplets with   three  equal variables,  the equality   of II$= H(X)$ holds, i.e.: II is equal to the Shannon entropy of the variable. The proof of this is  straightforward and   is similar to the one  derived for the mutual information limit.

\end{itemize}

\clearpage
\section*{Figure Legends}

\begin{figure}[!t]
\centering
\includegraphics[width=14cm]{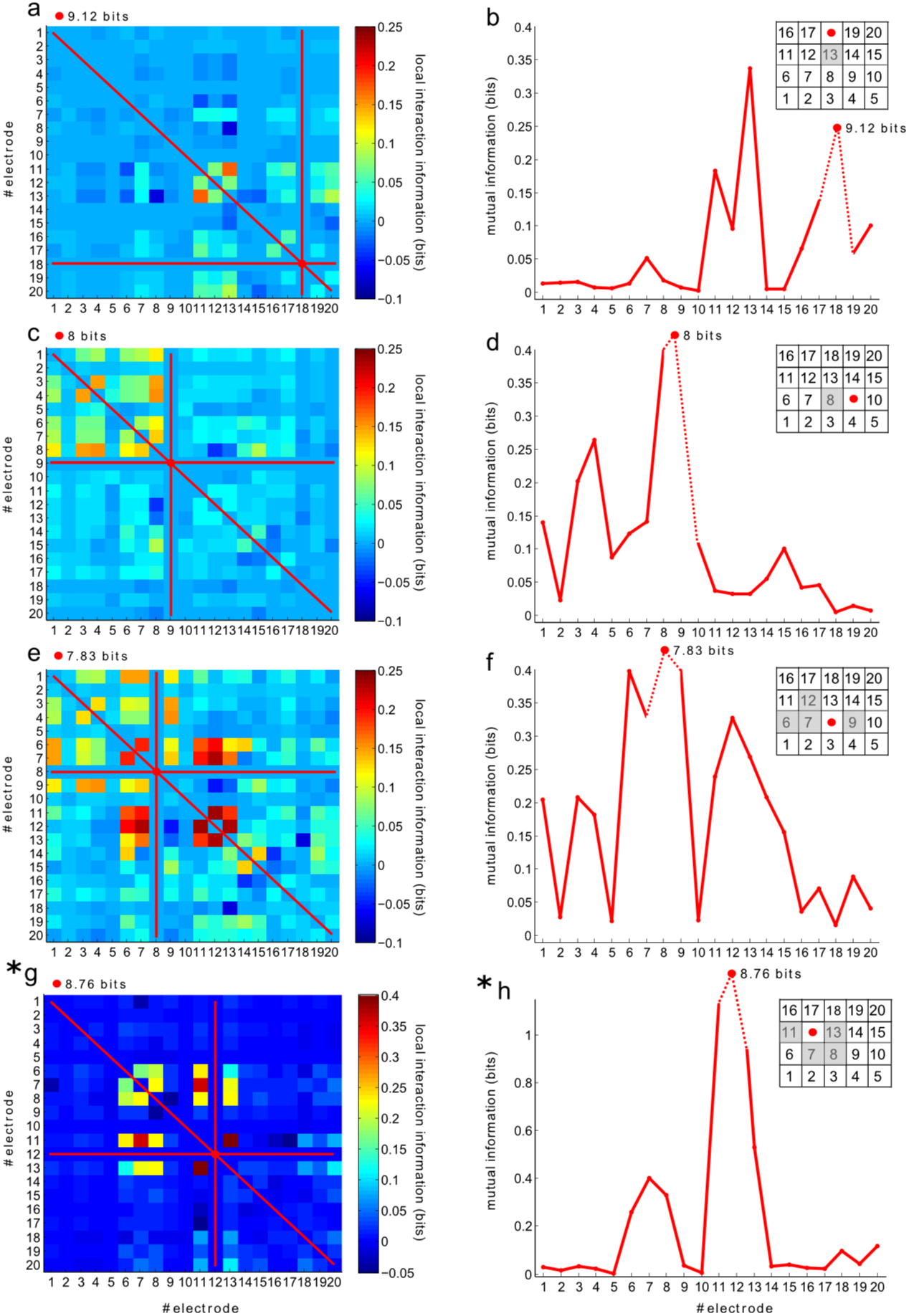}
\caption{}
 \label{fig1}
\end{figure}

\textbf{Figure \ref{fig1}: Structure of the interaction information in epilepsy based on human electrocorticography data.}   Positive values of II indicate R, whereas negative values indicate S. The 2D plots of II were obtained \JMC{by} varying two variables but keeping one of them fixed. In particular, we fixed the electrode with  \textbf{a:}  maximum entropy (\JMC{that for this particular case was} coincident with the electrode with maximum betweenness)  \textbf{c:}      minimum entropy  \textbf{e:}    maximum degree (the hub, \JMC{for this particular case was coincident} with the electrode with maximum clustering coefficient). \JMC{\textbf{g:}  }		   electrode number 12, \JMC{as a representative} belonging to the    resection area.  \textbf{a,c,e,g:}  red lines correspond to the mutual information between the electrode that was kept fixed and the rest of electrodes in the grid. Indeed, as shown in Appendix A, II is coincident with the mutual information when two variables are equivalent in the triplet. The solid red circle is the Shannon entropy value of the electrode that was fixed \JMC{(coincident with II(X,X,X)}).
\textbf{b,d,f,h:}   Same red lines as in panels a, c, e, g. Dashed lines have been introduced to show when the Shannon entropy value is much bigger than the values of mutual information. Electrocorticography grids are also plotted to show that similar values of mutual information were clustered within a geometrical similar region of the grid (in gray we plot the values of mutual information larger than 0.3 bits). Note that while II can be either positive or negative, the mutual information is strictly positive (when it is zero, statistical independence is indicated).  \textbf{g,h:} (*) Different scale.

\textbf{Figure \ref{fig2}: Synergetic  and redundant interactions in epilepsy from human electrocorticography data.}  \textbf{a:} Varying $\bar{Z}$ (as explained in \JMC{the} methods), we can sum all the negative values of  $II(X,Y,\bar{Z})$  to obtain the individual S (red) and similarly, the positive values of  $II(X,Y,\bar{Z})$  to achieve  the  individual R (blue). In green, we plot the Shannon entropy for each electrode (i.e.: $II(X,X,X)$,   see Appendix A for further details) and in magenta, the LSI as measured in  \cite{palmigiano2012},    summing the absolute value of the correlations between each electrode and its neighbouring electrodes in the grid. Note that for illustrative purposes the four measures are represented in an arbitrary scale. We plotted the maximum values for each measure in coloured circles.  \textbf{b:}  The six maximum electrode values for each of the measures plotted in panel a. In addition, we also plotted  the resection area \JMC{in gray}, in this case from a patient in which the seizures disappeared after surgery.

\textbf{Figure \ref{fig3}: Structure of the interaction information in epilepsy at different frequency-bands \JMC{obtained} from human electrocorticography data.}  As in figure \ref{fig1}, we  plot II  keeping one electrode belonging to the resection area fixed across the different frequency bands:  \textbf{a,b} delta (1-4 Hz) \textbf{c,d} theta (4-7 Hz) \textbf{e,f} alpha (7-14 Hz) \textbf{g,h} beta (14-26 Hz) and \textbf{i,j} gamma (26-100 Hz). One can see how the values of II  are higher at  $\beta$ and $\gamma$ bands. \textbf{b,d,f,h,j:}    \JMC{ Similar to figure \ref{fig1},  we plot in gray the values of mutual information larger than 0.3 bits. }

\textbf{Figure \ref{fig4}:  Redundancy and synergy  across frequency bands.}  \textbf{a:} R \JMC{receives a larger}  contribution from the high-frequency beta and gamma bands (orange and brown), with both bands showing a similar spatial grid map  \JMC{overlapping}  with the epileptogenic zone (cf. figure \ref{fig2}).  \textbf{b:} The low delta and theta \JMC{made a stronger} contribution to S (dark and light blue). Importantly, the spatial distribution of S is quite different from \JMC{that of} R: whilst the activation of R   \JMC{overlaps} the epileptogenic zone and it is represented by high frequencies, \JMC{S tends to surround  this region and it operates in low frequency bands.}

\textbf{Figure \ref{fig5}:  Validation across patients.}  \textbf{a:} We plotted the average values of R for each electrode inside the resection area (blue dashed line) and S outside the resection area (red solid line)  -- the total amount of R inside and S outside divided by the number electrodes in the resection area --.
This was done for 3 patients with Engel I (where the resection area matched or exceeded the epileptogenic zone) and for 1 patient with Engel III   (to illustrate at least one case where the resection area did not match the epileptogenic zone). The Engel I and III patterns were quite different. \textbf{b:}   \JMC{The ratio between R inside the resection area --R(in)-- divided by that of S outside that area --S(out)--, and both the ratios of S inside --S(in)-- divided by S(out) or R(in) divided by R(out).}  There are clearly important differences between Engel I and Engel III patients, corroborating the dynamic pattern found: R within the epileptogenic zone and S organized in the surrounding region. The total R for each patient is also represented \JMC{ (i.e., numbers 373.3, 665.2, 558.2 and 1296.1)} whereby the Engel III value is much larger than that for the Engel I cases.

\textbf{Figure \ref{fig6}: Comparison between the Gaussian   and the non-Gaussian assumptions.}  Panels a, c and e have been  plotted again from figure \ref{fig1} for ready comparison. The interaction information obtained when keeping the electrode fixed with:  \textbf{a,b:}   maximum entropy;  \textbf{c,d:}   minimum entropy \textbf{e,f:}   maximum degree (the hub \JMC{that in this particular situation} is coincident with the electrode with the highest clustering coefficient). \textbf{b, d, f:} The scenario for non-Gaussian data does not change qualitatively with respect to  the Gaussian assumption.
\textbf{g:}  As in figure \ref{fig2} but assuming   a non-Gaussian distribution. Note that the same pattern is observed for R whilst some differences are found \JMC{for} S. \JMC{Now, the statement of R and LSI having  a similar spatial grid, and S surrounding  R, is more clear than for the Gaussian case. }

\clearpage

\section*{Legends to Supporting Information}

\textbf {Figure \ref{figS1}: Sampling validation of the interaction information.}  The figure shows the coefficient of variation of the II (standard deviation of the estimation divided by its mean)   resulting from averaging over 9 different windows   and consisting of 2000 points each. One can see how the standard deviation is well-sampled and significantly smaller than the average.

\begin{figure}[!t]
\centering
\includegraphics[width=14cm]{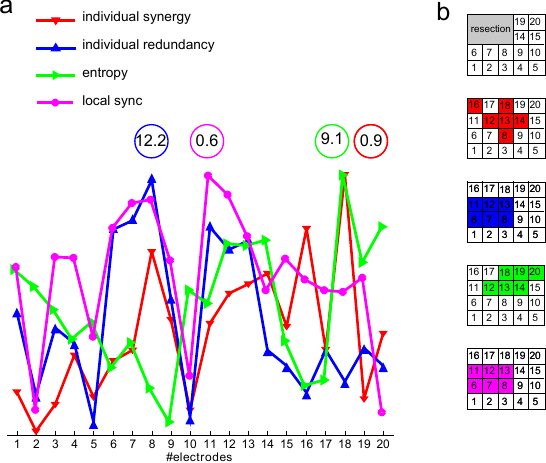}
\caption{}
\label{fig2}
\end{figure}

\begin{figure}[!t]
\centering
\includegraphics[width=12cm]{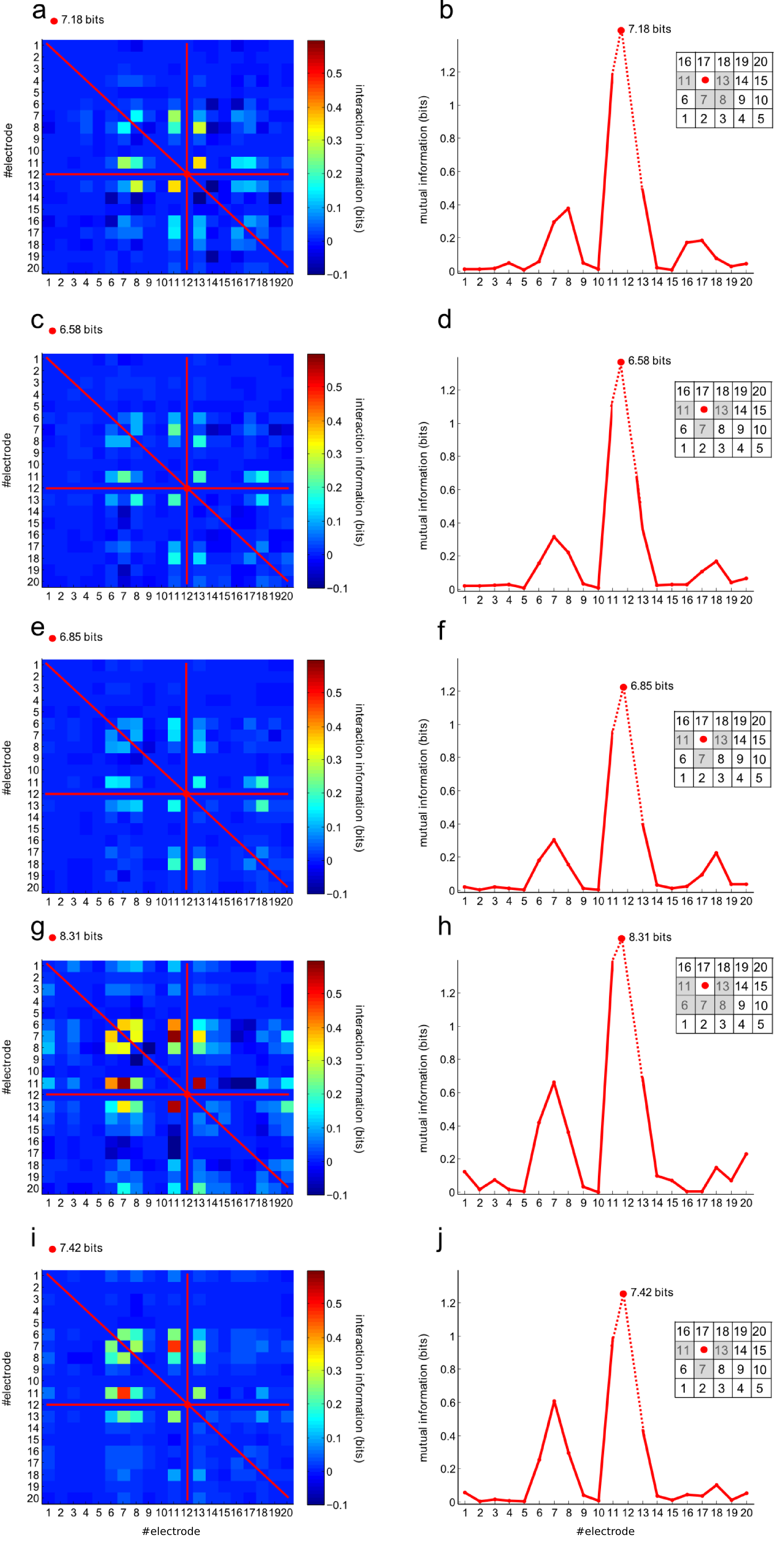}
\caption{}
\label{fig3}
\end{figure}

\begin{figure}[!t]
\centering
\includegraphics[width=14cm]{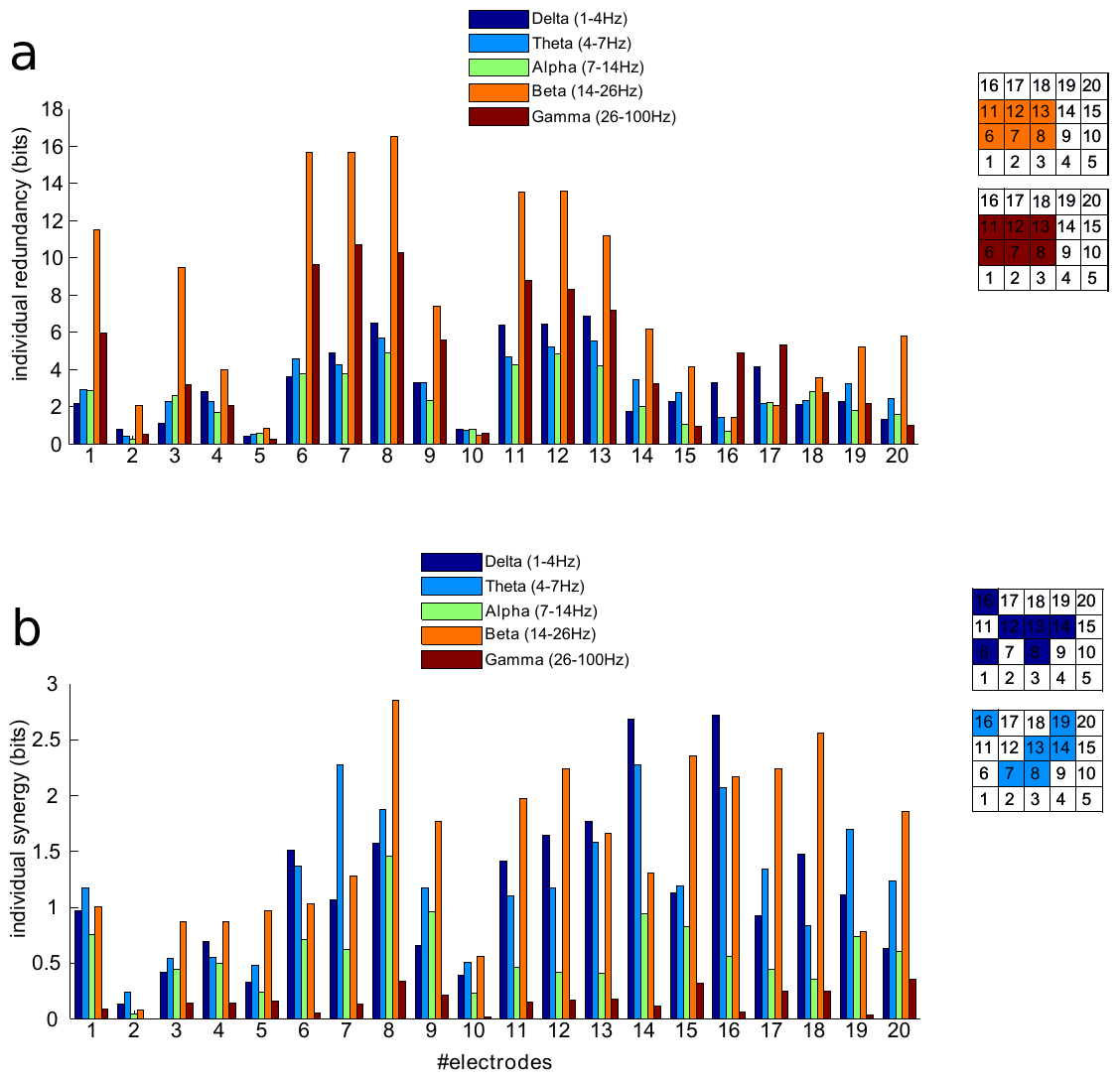}
\caption{}
\label{fig4}
\end{figure}

\begin{figure}[!t]
\centering
\includegraphics[width=14cm]{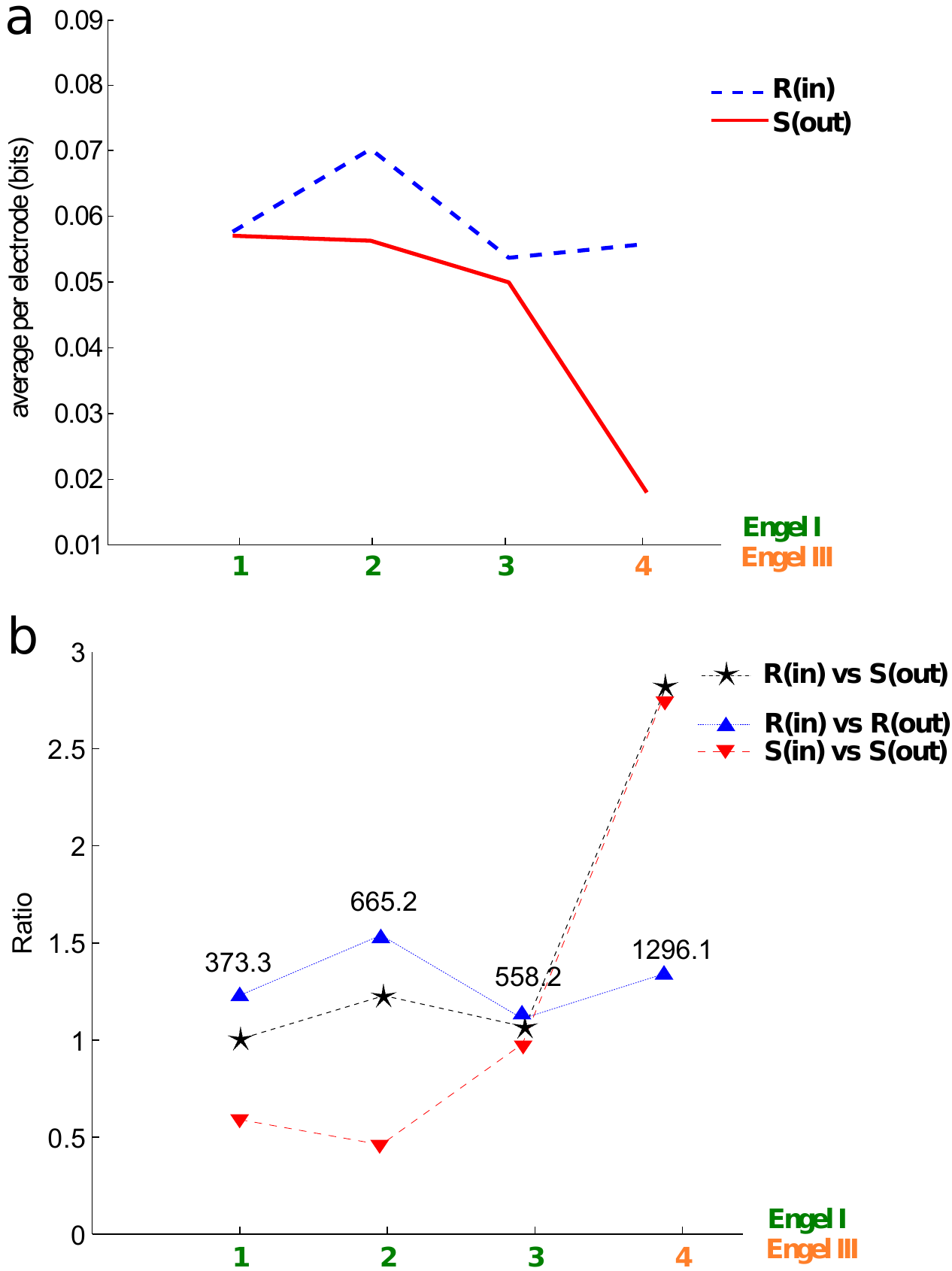}
\caption{}
\label{fig5}
\end{figure}

\begin{figure}[!t]
\centering
\includegraphics[width=11cm]{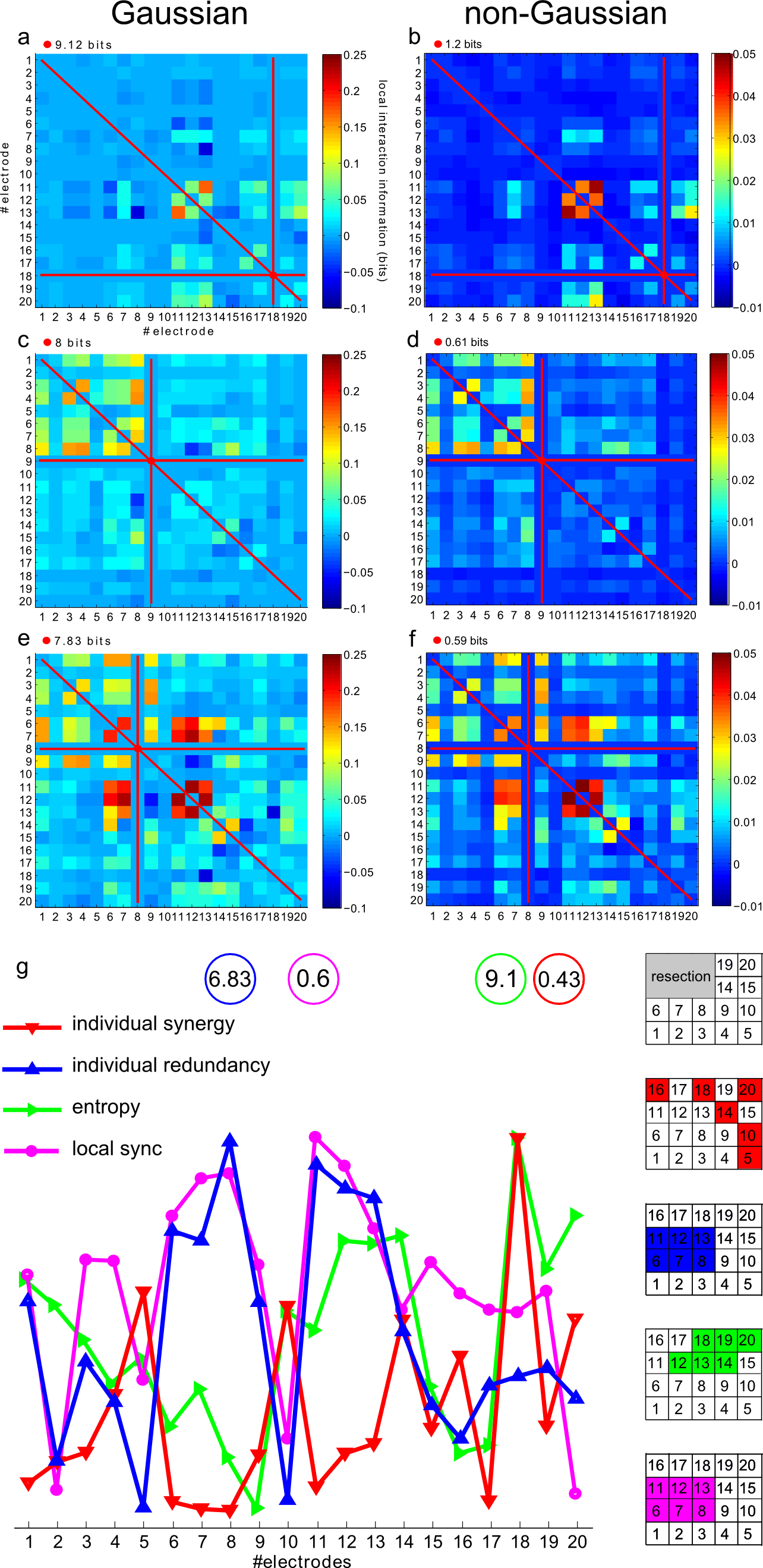}
\caption{}
\label{fig6}
\end{figure}

\JMC{\textbf {Figure \ref{figS2}: Redundant interactions in epilepsy from human electrocorticography  data for a patient with  Engel III.}  Similar to      figure \ref{fig2}  but for a patient with Engel III, i.e.,  with  patient's seizures not disappearing  after surgery. Observe that  the three electrodes 3, 8 and 13  have    maximum   LSI (local sync, an index which was proven to match   the epileptogenic zone in \cite{palmigiano2012})  and also   have  maximum R.     }

\JMC{\textbf {Figure \ref{figS3}: Evidence of non-Gaussianity in the data.}  Normal probability plots indicate strong data deviation from Gaussianity (the more the data do not follow the straight line, the more non-Gaussian is). \textbf{a:} data corresponding to one subject, one electrode and one of the segments used to average and    calculate II (and therefore, R and S).  \textbf{b:} Same data that in a, but binned to 4 states. Non-gaussianity also applied to binned data. \textbf{c,d:} Similar to panels a and b, for the same subject but different electrode. The same results are valid across subjects. }

\clearpage

\clearpage

\clearpage

\clearpage

\clearpage

\clearpage

\clearpage
\setcounter{figure}{0}
\renewcommand{\thefigure}{S\arabic{figure}}
\begin{figure}[h]
\centering{
\includegraphics[width=14cm]{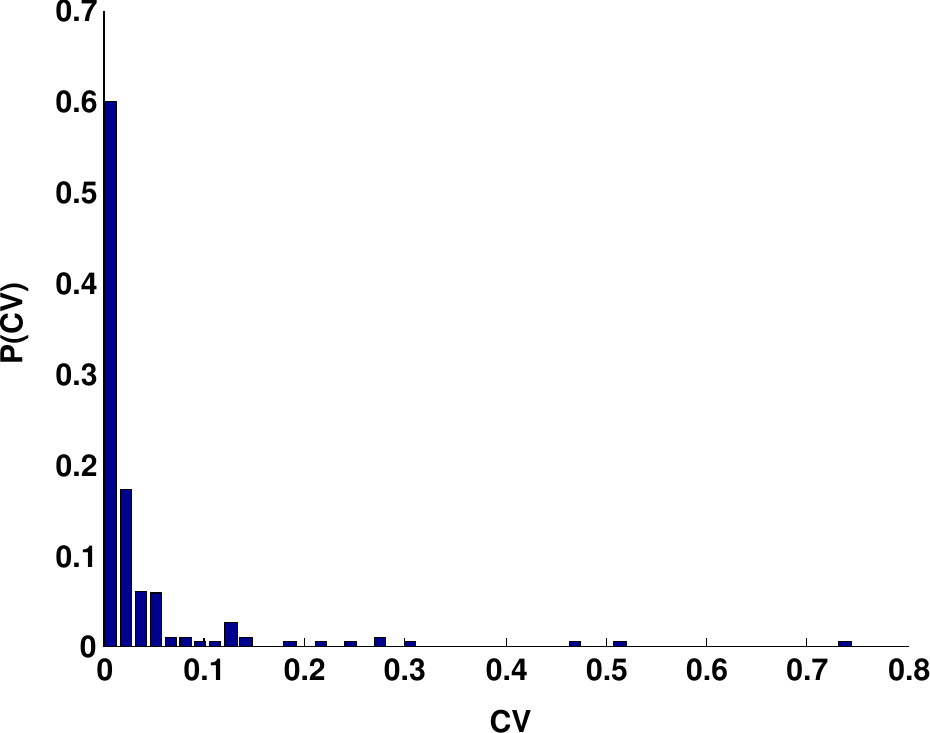}
}
\caption{}
 \label{figS1}
\end{figure}

\clearpage
\begin{figure}[h]
\centering{
\includegraphics[width=14cm ]{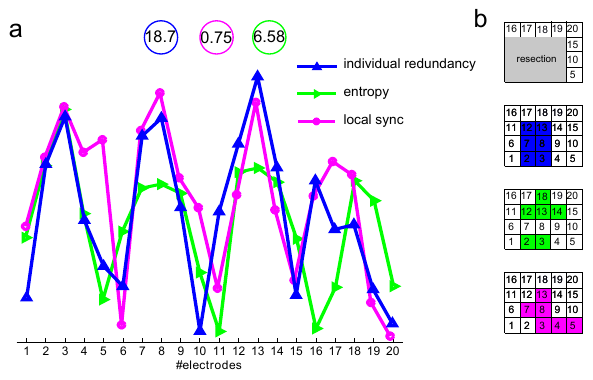}
}
\caption{}
 \label{figS2}
\end{figure}

\clearpage
\begin{figure}[h]
\centering{
\includegraphics[width=14cm ]{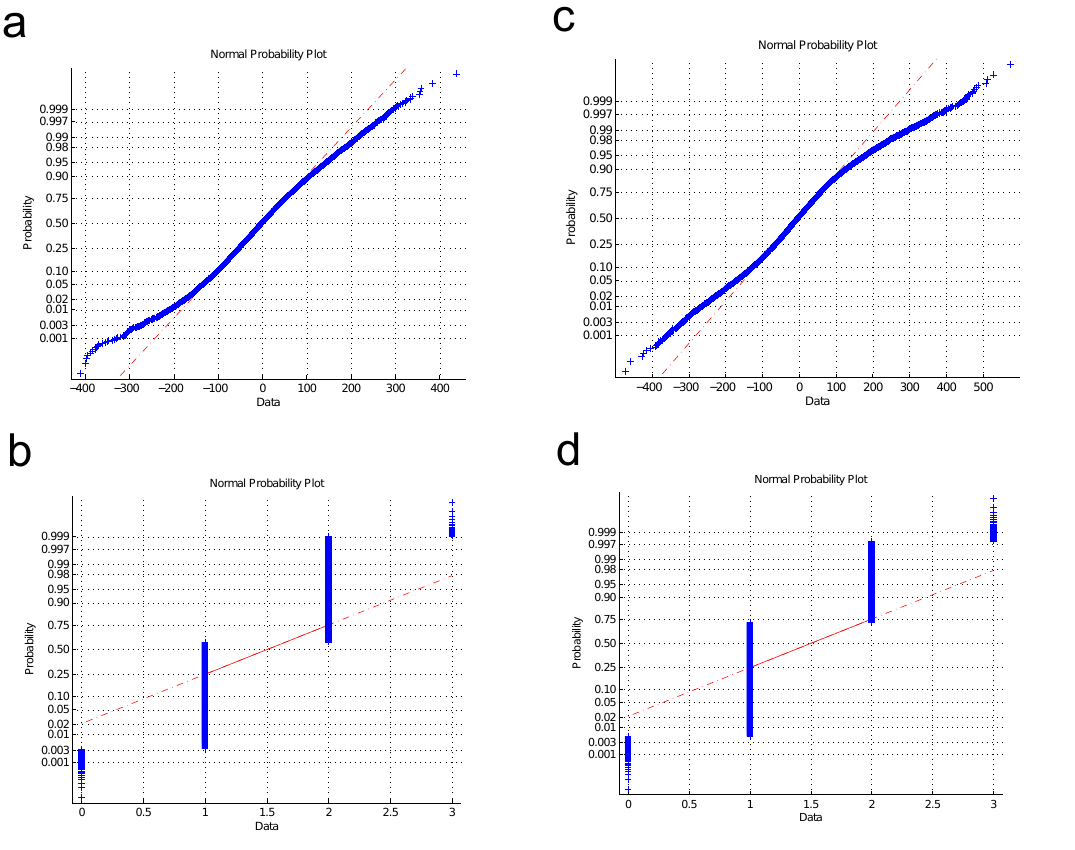}
}
\caption{}
 \label{figS3}
\end{figure}

\end{document}